\documentclass[prd,tightenlines,nofootinbib,amsfonts,amssymb,amsmath,color,11pt]{revtex4}
\usepackage{graphicx}
\usepackage{caption}
\usepackage{color}

\usepackage{amsfonts}
\usepackage{amsmath}
\usepackage{amssymb}
\usepackage{hyperref}

\def\be{\begin{equation}} 
\def\ee{\end{equation}}
\def\bea{\begin{eqnarray}}
\def\eea{\end{eqnarray}}

\begin{document}

\title{Anisotropic Inflation  with Derivative Couplings }
\author{Jonathan Holland$^{a}$, Sugumi Kanno$^{b,c}$, Ivonne Zavala$^a$}

\affiliation{$^a$  Department of Physics, Swansea University, Swansea, SA2 8PP, UK \\ 
$^b$ Department of Theoretical Physics and History of Science, University of the Basque Country
48080 Bilbao, Spain \\ 
$^c$ IKERBASQUE, Basque Foundation for Science, 
Maria Diaz de Haro 3,
Bilbao, 48013, Spain\\
}


\begin{abstract}
	
We study anisotropic power-law inflationary solutions when the inflaton and its derivative couple to a vector field. This type of coupling is motivated by D-brane inflationary models, in which the inflaton, and a vector field living on the D-brane, couple disformally (derivatively). We start by studying a phenomenological model where we show the existence of anisotropic solutions and demonstrate their stability via a dynamical system analysis. Compared to the case without a derivative coupling, the anisotropy is reduced and thus can be made consistent with current limits, while the value of the slow-roll parameter remains almost unchanged.  We  also discuss solutions for more general cases, including D-brane like couplings. 
 
\end{abstract}

 \maketitle

 \tableofcontents

 \section{Introduction}

The most recent data on the cosmic microwave background radiation (CMB) temperature fluctuations,  \cite{Planck15Infla,BICEP15}, favour  the simplest inflationary scenario to provide their origin. This simple scenario makes robust predictions for the primordial inhomogeneities of the universe:  they  are  adiabatic, highly isotropic,  approximately  scale-invariant,  and  nearly  Gaussian. In this framework, a single scalar field drives a period of quasi-de-Sitter accelerated expansion, while  its  quantum fluctuations are  stretched to observable scales, setting up the initial conditions for structure growth.  
Most inflationary models are based on scalar field dynamics, where inflation is driven by a single scalar field. 
The current data have severely constrained some of these models, however many  remain compatible  \cite{Vennin}. With  more  precise  observations  to  come  in the  future, it will be possible to determine whether more complicated models of inflation are needed,  which will narrow down the landscape of observably viable scenarios. 

In particular,  multifield models of inflation arise naturally in extensions of the simplest cosmological framework, for example in fundamental theories such as supergravity and string theory. Usually, models with more than one scalar field are considered in these extensions. However,  fields with other spins, such as gauge fields, may play an interesting and testable role  during inflation as well \cite{Soda,MSJS,Emami}.
Gauge fields are not commonly  considered  in the study of inflation, due to the cosmic no-hair conjecture, 
which states that spacetime rapidly approaches quasi-de-Sitter spacetime during inflation.
Thus, matter fields satisfying the dominant and strong energy conditions will be rapidly diluted.
However, \cite{WKS1} found the first working model (free from ghosts) of inflation with a vector field that can produce persistent anisotropy in the background spacetime. 
This brought forward the  interesting possibility that light gauge fields may affect cosmological observations by generating some observable amount of statistical anisotropy (for reviews on anisotropic inflation see \cite{Soda,MSJS,Emami}).  

Another possibility for generating observable statistical anisotropies in the presence of vector fields is the vector curvaton scenario \cite{Dimopoulos1,DKLR,DKW}. In this scenario, the inflaton is a scalar driving inflation, while  the vector field  becomes important after inflation, when it may dominate the universe and imprint its perturbation spectrum before it decays, as in the scalar curvaton scenario (for a review of the vector curvaton see \cite{Dimopoulos2}).
 
In the context of D-brane inflationary models, the inflaton is typically identified 
with the scalar field parameterising the transverse fluctuations of the D-brane (that is, its position in the internal compact six-dimensional space).  Such  a  brane  features  a  world-volume  two  form  field $F_{\mu\nu}$, associated with the longitudinal fluctuations of the D-brane. Therefore, it is natural to investigate the role of this brane field in the dynamics of inflation. Indeed, in \cite{DWZ}, a D-brane vector curvaton realisation was discussed, while in \cite{AZ} a  Wilson line inflationary model was studied with interesting predictions. In this case, it is precisely one of the D-brane vector internal components which drives inflation. 

In a D-brane scenario, the scalar field associated with brane position, identified with the inflaton, and the vector field couple disformally via the DBI action describing the D-brane dynamics\footnote{For an example of a theory where two scalar fields couple disformally via the DBI action, see \cite{DisformInf}.}. In particular, the gauge kinetic function, $f$, depends on the scalar field $\phi$ and its derivative $X\equiv \frac12(\partial \phi)^2$, $f(\phi,X)$. This coupling can thus alter the predictions of the anisotropic background evolution and  the predictions for statistical anisotropies.
Furthermore, $f(\phi,X)$ represents a general parametrisation of a generic inflaton-matter coupling. This has recently been  used in studies of the inflationary universe as a cosmological collider \cite{Chen}. It is thus important to know what effect the derivative coupling has on inflationary evolution. 
This derivative coupling  has also appeared in the recently proposed EFT of anisotropic inflation  \cite{Hassan,Rostami}.

Motivated by the D-brane scenario and the more generic nature of a derivative coupling between the inflaton and a vector field (and even more generally, with matter), in this paper we study   anisotropic inflation with derivative couplings. We start by considering a phenomenological model where the gauge kinetic function has a monomial  dependence on $X$ and exponential dependence on $\phi$. For the power-law cases we consider, no stable solutions exist for when $f(X)$ only. On the other hand, for $f(\phi,X)$, stable anisotropic solutions exist and the anisotropy is considerably reduced in comparison to the non-derivative case. This is interesting in view of the latest constraints on anisotropy \cite{KK,NKY}. We next use our general equations to explore more general solutions. Finally we conclude in section \ref{conc} with a discussion of our results and prospects for future work.

\section{Scalar-vector-tensor action with  general derivative couplings }\label{sec:1}

Although we will only look at power-law inflationary solutions, in this section we will keep the discussion as general as possible when presenting the set-up and equations of motion. This will be useful to describe a variety of power-law models as we discuss later. In most of the paper we will concentrate on a simple field theory model that will serve to illustrate the consequences of taking derivative couplings into account. 

Our starting point is the general scalar-vector-tensor action of the form:
\be\label{action}
S= \int{d^4x \sqrt{-g} \left\{\frac{M_{Pl}^2}{2} R - P(\phi, X) - \frac{f^{2}(\phi,X)}{4}  F_{\mu\nu}F^{\mu\nu}  \right\}}\,,
\ee
where $F_{\mu\nu} = \partial_\mu A_\nu - \partial_\nu A_\mu$,  $2X=(\partial\phi)^2$, thus we see that the gauge kinetic function depends both on the inflaton, $\phi$ and its derivative, $X$. This action is motivated from D-brane actions in string theory models of inflation, where $P(\phi,X)$ and $f(\phi,X)$ take very specific forms and arise from the Dirac-Born-Infeld (DBI) action (see \cite{DWZ} for details). Here we keep these functions general, in order to cover other possibilities\footnote{Anisotropic inflationary solutions with a DBI kinetic term for the scalar field  and a pure inflaton-dependent gauge kinetic function $f$ were considered in \cite{DoKao}. More general forms for $P(\phi, X)$ were further considered in \cite{OST}. }. 

The equations of motion derived from \eqref{action} are given by 
\bea
&&R_{\mu\nu} -\frac{1}{2} g_{\mu\nu} R = 8\pi G \left(T^A_{\mu\nu} + T^{\phi}_{\mu\nu} \right) \,, \\
&& \frac{1}{\sqrt{-g}} \partial_\mu \left[ \sqrt{-g} \left( \frac{F^2}{2} f f_X  + P_X \right) \partial^\mu \phi \right] = \frac{F^2}{2} f f_\phi + P_\phi \,,\\
&&  \partial_\mu \left[ \sqrt{-g} \,f^2(\phi,X) F^{\mu\nu} \right]  =0\,, 
\eea

\noindent where $8\pi G = M_{Pl}^{-2}$ is the reduced Planck mass and we have denoted the derivatives as $f_i = \partial_i f$ and similarly for $P$, for $i=\phi, X$. 
The energy-momentum tensors for the vector and the scalar fields are given by 
\bea
&& T^\phi_{\mu\nu}= \partial_\mu \phi \,\partial_\nu \phi \left(\frac12 f f_X F^2 + P_X \right)  -g_{\mu\nu} P  \,,\\
&&  T^A_{\mu\nu} = f^2 \left[F_\nu^{\,\,\alpha} F_{\mu\alpha} - g_{\mu\nu}  \frac{F^2}{4}\right] \,.
\eea

 We are interested in anisotropic solutions and therefore, without loss of generality,  we consider the following anisotropic metric: 
  \be
  ds^2 = -dt^2 + e^{2\alpha(t)} \left[ e^{-4\sigma(t)} dx^2 + e^{2\sigma(t)} (dy^2 + dz^2)\right]\,
  \ee
 where $e^{\alpha(t)}$ is identified with the isotropic scale factor, and $e^{\sigma(t)}$ characterises the anisotropy.
Furthermore, we use gauge invariance to choose $A_0=0$ and, for concreteness, we consider homogeneous fields of the form  \cite{WKS1,KSW}: 
\be
\phi = \phi(t)\,, \qquad  A_\mu = (0,v(t),0,0)\,.
\ee 
 With these Ans\"atze, the equation of motion for the vector field takes the simple form:
 \be
 \frac{d}{dt} \left[ f^2 e^{\alpha + 4\sigma} \dot v \right] =0\,,
 \ee 
  which can be readily solved to give: 
  \be\label{fsol}
  f^2 e^{\alpha + 4\sigma} \dot v = p_A\,,
  \ee
  where $p_A$ is a constant of integration. Since $- \dot v=F_{x0}=E_x$, $p_A$ is  the electric field modulated by the expansion of the universe.
  
The Einstein equations, on the other hand, can be arranged into the following set of equations   
  \bea
  && \ddot \alpha = -3\,\dot \alpha^2 + \frac{1}{6M_{Pl}^2} \left[ 6P +  f^{2}{\dot v}^2 e^{-2\alpha+4\sigma} + 3\,\dot \phi^2 \left(
  P_X -  \frac{f_X}{f} f^{2}{\dot v}^2 e^{-2\alpha+4\sigma}   \right)\right]   \label{einstein1} \,, \\
  &&\ddot \sigma= -3\,\dot \alpha\, \dot \sigma + \frac{f^{2} {\dot v}^2 }{3M_{Pl}^2 } e^{-2\alpha+4\sigma} 
   \label{einstein2} \,, \\
  && \dot \alpha^2 = \dot \sigma^2 + \frac{1}{3M_{Pl}^2} \left[ P + \frac{f^{2}{\dot v}^2}{2} e^{-2\alpha+4\sigma} + \dot \phi^2 \left(
  P_X - \frac{f_X}{f} f^{2} {\dot v}^2  e^{-2\alpha+4\sigma}\right) \right]  \label{einstein3}\,, 
  \eea
  where we have used \eqref{fsol}. Finally, the equation of motion for the scalar field becomes:
 \bea\label{fieq}
 &&\hskip-0.5cm \ddot \phi \left[-\dot \phi^2 \,P_{XX}+ P_X - f^{2} {\dot v}^2  e^{-2\alpha+4\sigma} \left(\frac{f_X}{f}+3\frac{f_X^2}{f^2} \dot \phi^2 - \frac{f_{XX}}{f} \dot \phi^2 \right) \right]  \nonumber \\
&&\hskip 0.5cm
  + \, \dot \phi \left[  \dot\phi \, P_{X\phi} + 3\,\dot\alpha P_X   + f^{2} {\dot v}^2  e^{-2\alpha+4\sigma}  \, \frac{f_X}{f} \left(\dot \phi \left(3 \frac{f_{\phi}}{f}  - \frac{f_{X\phi}}{f_X} \right) +4 \dot \sigma + \dot \alpha \right) \right] 
 \nonumber \\
 && \hskip8cm 
 + P_\phi  -  \frac{f_\phi}{f} f^{2} {\dot v}^2 e^{-2\alpha+4\sigma}  =0\,.
 \eea
From these equations, it is easy to recover the various examples studied in the literature, for which $f_X=0$ \cite{WKS1,KSW,DoKao,OST,IS}.
 
 In what follows we use these equations to look for  stable anisotropic solutions. We start  by looking at a phenomenological example that serves as a prototype to understand the effect of the derivative coupling between the inflaton and the vector field, and then we explore more general cases.  
 

\section{Anisotropic power-law  inflation with derivative couplings}\label{Sec:3}

In this section we start the analysis of power-law  anisotropic inflation with derivative couplings,  providing  the first  explicit example of the situation described in the EFT description of  \cite{Hassan}. 
  We start with a canonically normalised inflaton:
  \be\label{Peq} 
  P(\phi, X) = \frac{1}{2} (\partial\phi)^2 + V(\phi)  = X+V\,, 
  \ee
  and thus  replace, $P_{\phi} = V_{\phi}$, $P_X=1$ in the equations of motion above, \eqref{einstein1}-\eqref{fieq}. 
Note that  eqs.~\eqref{einstein1} and \eqref{einstein3}  depend only on the derivative of $f$ w.r.t. $X$. One then immediately sees that a suitable choice of functional form is given by setting: 
\be
X\frac{f_X}{f} = -n  \,,
\ee
where $n = $ const. This has the solution: 
\be
f(X,\phi) = (-X)^{-n} g(\phi) \,,
\ee
for some function $g(\phi)$. On the other hand, we can also see that a suitable choice of   $\phi$ dependence is given when $f_\phi/f=\,$const. that is, an exponential dependence. So we find  that a suitable Ansatz for the gauge kinetic function's dependence on the scalar and its derivative is given by\footnote{In a string theory scenario, the effective 4D action can be written in terms of the 4D $M_{Pl}$, which would be a function of the string scale and coupling, as well as the compactification volume.}:
\be\label{gkf}
f(\phi,X) = M_{Pl}^{4n}\,f_0\,\frac{e^{\frac{\rho}{M_{Pl}} \, \phi}}{(-X)^n} \,.
\ee
In addition to this,  we also consider  an exponential potential for the scalar field
\be\label{Vansatz}
V(\phi) = V_0 \, e^{\frac{\lambda}{M_{Pl}} \phi}\,.
\ee
We are now ready to   look for power-law solutions of the form: 
\be\label{ansatz}
\alpha = \zeta \log (M_{Pl} \, t) \,, \qquad \sigma = \eta \log (M_{Pl} \, t) \,,  \qquad \frac{\phi}{M_{Pl}} = \xi \log (M_{Pl} \, t) + \phi_0   \,.
\ee
Using this Ansatz with the Hamiltonian constraint \eqref{einstein3}, we obtain the conditions:
\be\label{condition1}
\lambda\xi = -2  \,, \qquad \rho\,\xi + 2\,\zeta + 2\,\eta + 2n =1 \,. 
 \ee
We arrive at these two conditions \eqref{condition1} by requiring that, after substitution of \eqref{ansatz} into \eqref{einstein1}-\eqref{fieq}, powers in $t$ balance in all equations (i.e., we end up with equations of the form $Ct^n=Kt^n$ where $C,K$ are independent of $t$). The remaining conditions (below) come from ensuring that $C=K$ in \eqref{einstein1}-\eqref{fieq}, i.e. that the equations are satisfied (the amplitudes balance) after substitution of \eqref{ansatz}. For the amplitudes to balance in \eqref{einstein3} (the Hamiltonian constraint), it is required that:
 \be
 -\zeta^2 + \eta^2 + \frac{1}{6} \,\xi^2 + \frac{1}{3}\,u + (1-4n)\,4^{-n}\, \xi^{4n}\, \frac{w}{6}=0\,, \label{Ham Constraint}
 \ee
 where we have defined $u$, $w$ as:
 \be
 u = \frac{V_0}{M_{Pl}^4}\, e^{\lambda \phi_0} \,, \qquad w = \frac{p_A^2}{M_{Pl}^4}\,f_0^{-2}\,e^{-2\rho \phi_0} . \label{u w}
 \ee
 From the equation for the scale factor \eqref{einstein1} we then obtain:
 \be
3\zeta^2 -\zeta - u +4^{-n}\,(6n-1)\, \xi^{4n}\, \frac{w}{6} =0 \,, \label{Scale Factor}
 \ee
Similarly from the anisotropy equation \eqref{einstein2} we get: 
\be
 -\eta\, ( 3  \zeta -1) - 4^{-n}\, \xi^{4n}\, \frac{w}{3} =0\,. \label{Anisotropy}
\ee
Finally from the equation for the inflaton we obtain:
 \be
-\xi +3\,\zeta\,\xi+\lambda\,u+4^{-n}\,\xi^{4n-1}\,w\,\left[2\,n\,(\zeta+4\,\eta+4\,n-1)+\rho\,\xi\,(4\,n-1)\right]=0 \,.\label{Inflaton}
  \ee
 Using these equations, we can solve for $u$ and $w$ to get:
 \be\label{uw}
 u = -\zeta+3\,\zeta^2+\frac12\,\eta -3\,n\,\eta -\frac{3}{2}\,\zeta\,\eta+9\,n\,\zeta\,\eta , \quad
\qquad  w = 3\,\eta\,4^n\,\xi^{-4n}\,(3\,\zeta-1). 
 \ee
 Substituting these into the inflaton equation (\ref{Inflaton}), and using the constraints for $\xi$ and $\eta$ from (\ref{condition1}) gives:
 \bea
 && (-1+3\,\zeta)\left[8+\lambda^2 \,(1-6\,\zeta+2n\,(-1+9\,\zeta\,(-1+2n+2\zeta)))\right. \nonumber  \\ 
 && \hskip5cm  \left.-4\,\lambda\,\rho\,(-2+3\,\zeta+3\,n\,(1+3\,\zeta))+12\,\rho^2\right]=0\,.
\eea

In contrast with \cite{KSW}, we now obtain a cubic, rather than a quadratic  equation for $\zeta$. As in  \cite{KSW}, we have the solution $\zeta=\frac{1}{3}$ which gives $u=w=0$, implying that there is no anisotropy and no potential driving inflation. We hence discard this solution and focus on the other two:
 \be\label{zetas}
  \zeta_+=\frac{A+\sqrt{B}}{72\,n\,\lambda^2}  \,, \qquad \quad
   \zeta_-=\frac{A-\sqrt{B}}{72\,n\,\lambda^2}\,,
 \ee
 where
 \be
 A = 6\,\lambda^2+18\,n\,\lambda^2-36\,n^2\lambda^2+12\,\lambda\,\rho+36\,n\lambda\,\rho\,,
 \ee
 and
 \be
 \hskip-0.5cm B=A^2-144\,n\,\lambda^2\,(8+\lambda^2-2n\,\lambda^2 +8\,\lambda\,\rho-12\,n\lambda\,\rho+12\,\rho^2)\,.
 \ee
 These solutions trivially satisfy (\ref{Ham Constraint}), and it is important to remember  that they are constrained from the requirement that  $w, u$ must be positive by definition, (\ref{u w}).
 
To look for inflationary solutions we define the  average slow-roll parameter, $\epsilon$ in terms of the Hubble parameter defined by $H=\dot \alpha$, as:
 \be\label{eps}
 \epsilon \equiv -\frac{\dot{H}}{H^2}=\frac{1}{\zeta}  \,.
 \ee
 Hence there are two branches of  solutions  for  $\epsilon$ corresponding to $\zeta_\pm$. In order to have inflation, we  need $\epsilon \ll 1$, that is, we are looking for regions in the parameter space where $\zeta_{\pm}\gg 1$. 
 
 The anisotropy is  characterised by:
 \be\label{aniso}
 \frac{\Sigma}{H}\equiv \frac{\dot{\sigma}}{\dot{\alpha}}=\frac{\eta}{\zeta}\,,
 \ee
 where $\eta$ is given by:
 \be
 \eta = \frac{1}{2}+\frac{\rho}{\lambda}-\zeta -n\,.
 \ee
 As for $\epsilon$, there are two possible branches of solutions, associated to $\zeta_\pm$. 
 
Let us now discuss  two  cases of interest. Firstly, $\rho=0$, which corresponds to a gauge kinetic function that depends only on the derivative of the scalar field. And secondly, $\rho\ne 0$, when it depends on both. 
 
  \bigskip

\paragraph{Shift symmetric coupling, $\rho=0$. }

A purely shift symmetric coupling of the inflaton with the vector field arises for $\rho=0$ (see \eqref{gkf}). This type of coupling of the inflaton to matter was considered recently in \cite{Chen}.
 In order for inflationary solutions to arise, we need $\zeta\gg 1$.
 Moreover, the solutions should satisfy $u, w>0$ (see \eqref{uw}). 
 In terms of $\zeta$ (with $\rho=0$), $w$ and $u$ are given by:
\be
w=-3\times 2^{-2n-1} (3\zeta-1)\left(-\frac{1}{\lambda}\right)^{-4n} \left(2\zeta +2n -1\right) \,,
\ee
\be
u=-\frac14 (2n-1)(3\zeta-1)(6\zeta+6n-1) \,.
\ee
Therefore, in order for $w,u$ to be positive, $n$ must be negative and $|n| \gg 1$, so that  $n < 1/2 - \zeta$.  From the expressions for $\zeta_{\pm}$ \eqref{zetas} we see that for $|n| \gg 1$, $\zeta_\pm \sim  \frac{n}{2} (-1\pm1)$.\footnote{Notice that to see that $u$ is  positive in this limit (as required) one needs to include the next to leading order term in the large $n$ expansion for  $\zeta_{\pm}$.} Thus in principle there are anisotropic inflationary solutions that satisfy all necessary conditions for sufficiently large $|n|$. For example, $n=-10^5, \lambda =1$, gives $\epsilon \sim 10^{-4}, \Sigma/H \sim 10^{-9}$, as required by current bounds \cite{KK,NKY}. Note  that requiring sufficiently small  anisotropy requires very large values of $|n|$. However,  we find no stable solutions when $\rho = 0$.\footnote{This has been confirmed recently in \cite{DoK}.} 

\paragraph{More general coupling, $\rho\ne 0$.}
Let us now consider the case when the gauge kinetic function depends on both the inflaton and its derivative, that is $\rho\ne 0$.
As before, inflationary solutions require $\zeta\gg 1$. Furthermore, the conditions for a  positive $w$ (and $u$) can be obtained by looking at  \eqref{uw}, which takes the form:
\be
w=-3\times 2^{-2n} (3\zeta-1)\left(-\frac{1}{\lambda}\right)^{-4n} \left(\zeta +n -\frac{\rho}{\lambda} -\frac{1}{2}\right) \,,
\ee
Therefore for sufficiently large $\zeta$, $w$ can be positive for positive or negative $n$ and  large values of $\rho/\lambda$, which is also required to obtain large values of $\zeta$ (see eq.~\eqref{zetas}). In the limit  $\rho/\lambda\gg 1$, the solutions $\zeta_\pm$ become:
\be
\zeta_{\pm}\simeq\frac{\rho\left(1+3n\pm\sqrt{(3n-1)^2-\frac{8n}{\rho^2}}\,\right)}{6n\lambda} \, .
\ee
Examining this, we see that for $n>0$, the numerator is always positive since $(1+3n)>\sqrt{(3n-1)^2-(8n)/\rho^2}$ meaning $\zeta_{\pm}$ are both positive. Similarly for $n<0$, we see that $|1+3n|<\sqrt{(3n-1)^2-(8n)/\rho^2}$ which tells us that $\zeta_+$ is negative while $\zeta_-$ is positive. That is, in this limit, there are positive solutions for both $\zeta_\pm$ for  positive $n$, while for negative $n$ only one solution is positive. We are also interested in small anisotropy, $\Sigma/H\ll1$ \eqref{aniso}, where in terms of the parameters we have:
\be
\frac{\Sigma}{H} = \frac{1}{\zeta} \left(\frac{\rho}{\lambda} +\frac{1}{2} -n-\zeta \right)\,,
\ee 
which   further selects the  appropriate solution for $\zeta$\footnote{We focus only on solutions with $|n|\geq1$, which guarantees real values for $w$ in the $\lambda>0$ we are interested in.}. 
With these conditions, one can check that there is a range of values for which anisotropic solutions exists for $n>0$. A stability analysis shows that for these cases, both the isotropic and anisotropic solutions are attractors. Therefore the evolution of the system depends on the initial conditions. Since we are interested in the case where the anisotropic solution is the only attractor, in what follows we focus our search for anisotropic stable solutions to the case $n<0$.

\bigskip 

\begin{figure}
\centering
  \centering
  \includegraphics[width=\linewidth]{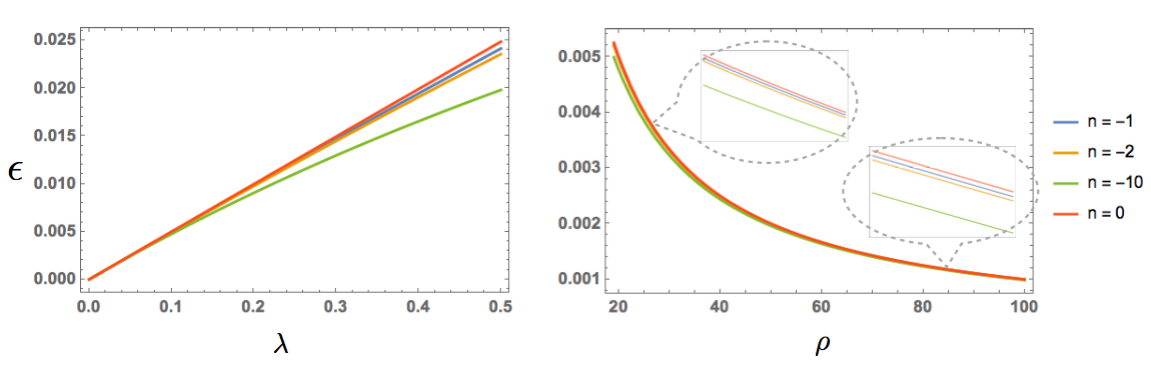}
  \captionof{figure}{In these plots we show how the slow-roll parameter, $\epsilon$, varies with $\lambda$ (for $ \rho=20$) and $\rho$  (for $\lambda=0.1$), for the values of $n$ shown.}
  \label{figeps}
\end{figure}

\begin{figure}
\centering
  \centering
  \includegraphics[width=\linewidth]{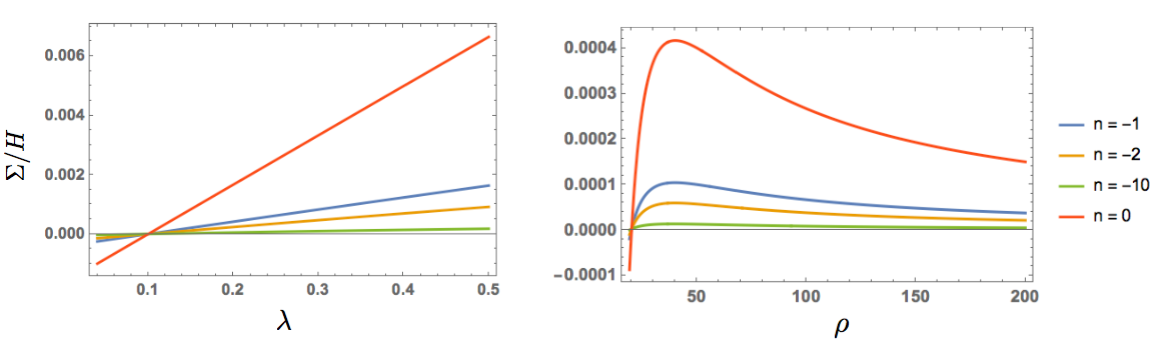}
  \captionof{figure}{In these plots we show  how the anisotropy, $\frac \Sigma H$, varies with $\lambda$ (for $\rho=20$) and $\rho$ (for $\lambda=0.1$), for negative values of $n$ as shown.}
  \label{figani}
\end{figure}

In Figure \ref{figeps}  we show the behaviour of  the slow-roll parameter, $\epsilon$, as a function of the  parameters $\lambda$ and $\rho$ for different values of $n$. As one can see, the slow-roll parameter $\epsilon$ decreases very slightly as the magnitude of $n$ increases. Conversely, as can be seen in Figure \ref{figani} the  anisotropy, $\Sigma/H$, can be reduced by the introduction of a derivative coupling: the greater the magnitude of $n$, the smaller the magnitude of anisotropy. 

We can understand the decrease in the anisotropy as follows. Since we are only interested in   solutions where the anisotropic point is a single attractor (and since the only observable anisotropic effects come from the final value of the anisotropy), we do not have to worry about the initial value for the gauge field.  If the  anisotropy converges to a number, its  final value is given by the ratio of energy density of the vector field to that of the scalar field  \cite{WKS2}. From eq.~\eqref{einstein2}, we can define this ratio $\cal R$ as:
 \be\label{ratio1}
 \frac{\Sigma}{H}  \simeq \frac{2}{3} {\cal R} \,, \qquad \,\,\,{\cal R} = \frac{\rho_v}{V(\phi)} \sim \frac{\frac{f^2}{2} \dot v^2 e^{-2\alpha+4\sigma}}{V(\phi)}\,.
 \ee
Using  \eqref{fsol}, \eqref{gkf} and \eqref{Vansatz}, this ratio can be written as 
\be
{\cal R} \sim \dot \phi^{4n} e^{-2\rho\phi-4\alpha-4\sigma -\lambda\phi} \sim \xi^{4n} t^{-2\rho\xi-4\zeta-4\eta -\lambda\xi -4n}\,,
\ee
where in the second expression we used \eqref{ansatz}. Furthermore, using the conditions \eqref{condition1}, we find that the ratio becomes a constant given by 
\be\label{ratio}
{\cal R} \sim \xi^{4n} = \frac{1}{\xi^{4|n|}}\,,      \quad   n<0\,,
\ee
and thus we see why the  anisotropy decreases with $ |n|$  in the  case with derivative couplings. Since  $\xi^{4n}$ is  coming from the energy density of the vector field, we see that the anisotropy is reduced because the energy density of the vector field becomes small during inflation.

\bigskip 

 \subsection{Stability of the anisotropic solutions }

We now study the stability of the solutions above using a dynamical system analysis.  
For this, we define the dimensionless variables:
\be\label{wyz}
 W=\frac{\dot{\sigma}}{\dot{\alpha}}, \, \qquad Y=\frac{1}{M_{Pl}}\,\frac{\dot{\phi}}{\dot{\alpha}},\, \qquad Z=\frac{f(\phi,X) \,e^{-\alpha+2\,\sigma}}{M_{Pl}}\,\frac{\dot{v}}{\dot{\alpha}},
 \ee
 where we use the e-fold number as time coordinate,  $d\alpha=\dot{\alpha}\,dt $. 
Using these variables, the Hamiltonian constraint  \eqref{einstein3} can be written as
\be\label{Hami2}
-\frac{V}{M_{Pl}^2 \dot\alpha^2} =  3(W^2-1)+\frac{Y^2}{2}+\frac{Z^2}{2}\left(1-4n \right)\,.
\ee
Since the inflationary potential is positive definite, we immediately see that \eqref{Hami2} implies 
\be
W^2 +\frac{Y^2}{6}+\frac{Z^2}{6}\left(1-4n\right) <1\,.
\ee
In terms of the variables \eqref{wyz}, the slow-roll parameter $\epsilon$, becomes:
\be
\epsilon=3W^2+\frac{Y^2}{2}+\frac{Z^2}{3}\left(1-3n\right). \label{eps3}
\ee
 Using the Hamiltonian constraint \eqref{Hami2}, the equations of motion in terms of \eqref{wyz} can be written as:
 \be
 \frac{dW}{d\alpha} =\frac{1}{3}Z^2\,(W+1-3 \,n\,W)+W\left(3(W^2-1)+\frac{1}{2}\,Y^2\right) \,,\label{Weq2} 
 \ee
 \bea
 &&\hskip-1cm\frac{dY}{d\alpha}=\frac16Y\left\{18W^2+3Y^2+2Z^2-6n\,Z^2+3C(Y,Z)\, \left[-4n\,Z^2\,(1+4W)  \right. \right. \nonumber \\
 &&\left. \left. +Y\left(-6Y+6\lambda\,(W^2-1)+\lambda\,Y^2-Z^2\,(-1+4n)(\lambda+2\rho)\right)\right] \right\} \,,\label{Yeq2} 
\eea
\bea
 &&\hskip-1cm\frac{dZ}{d\alpha}=\frac16Z\left\{-12-12W+18W^2+3Y^2+2Z^2-6nZ^2-6\rho Y -6nC(Y,Z)\, \left[4nZ^2\,(1+4W) \right. \right. \nonumber \\
 && \left. \left. -Y(-6Y+6\lambda(W^2-1)+\lambda Y^2-Z^2\,(4n-1)(2\rho+\lambda))\right]\right\} \,,\label{Zeq2}
\eea 
where
\be
C(Y,Z)=\frac{1}{Y^2+2nZ^2\,(1-4n)}\,.
\ee
We can now  find the fixed points of the system by setting  $ dW/d\alpha=dY/d\alpha=dZ/d\alpha=0$. From \eqref{Weq2}, we find:
\be
Z^2 =\frac{3W\,(-6+6W^2+Y^2)}{2(-1-W+3nW)} ,
\ee
These equations are solved numerically for suitable values of the parameters $\lambda$, $\rho$, and $n$ (chosen so that $w$ is positive) such that  $W$, $Y$ and $Z$ are all non-zero and real.
To choose appropriate solutions, we perform a linear stability analysis. 
The isotropic fixed point solution is located at $W=Z=0$, $Y=-\lambda$,  corresponding the coupling $f(\phi,X)$ being switched off. The linearised equations of motion around this point reduce to:
\bea
&& \frac{d \delta W}{d \alpha} = \left(\frac12 \lambda^2-3\right) \, \delta W \\
&& \frac{d \delta Y}{d \alpha} = \left(\frac12 \lambda^2 -3\right)\, \delta Y \\
&& \frac{d \delta Z}{d \alpha} = \left[ \left(\frac12 -n\right)\,\lambda^2 + \rho \, \lambda -2 \right]\,\delta Z \, .
\eea
When $\lambda$ is small, the LHS of these equations are all negative (corresponding to the isotropic fixed point being an attractor solution) if $\lambda^2 \,(1-2n)+2 \, \rho\,\lambda < 4$. If however $\lambda^2 \,(1-2n)+2 \, \rho\,\lambda > 4$, the isotropic fixed point is unstable. Since we are searching for anisotropic solutions, this is the parameter space we want to consider: we require that $ \rho > \frac2\lambda -\frac\lambda2\,(1-2n)$.

Now, we look at two explicit examples  to demonstrate  that  stable derivative anisotropic solutions can be found with small but non-zero anisotropy in agreement with recent data. 
Consider first the case with  $n=-1$, $\lambda = 0.1$, and $\rho = 20$, which  has  a fixed point at $(W,Y,Z)=(3.08249 \times 10^{-6},-9.92559 \times 10^{-2},\pm 5.26275 \times 10^{-3})$. Linearisation around this point gives:
\bea
&&\frac{d\delta W}{d\alpha}=-2.99504\,\delta W-3.05955\times 10^{-7} \,\delta Y \pm 3.50854 \times 10^{-3}\,\delta Z  \\
&&\frac{d\delta Y}{d\alpha}=-2.29684 \times 10^{-3}\,\delta W - 3.08768\,\delta Y \pm 0.87150 \, \delta Z \\
&&\frac{d\delta Z}{d\alpha}=\mp1.07688 \times 10^{-2}\,\delta W \mp 0.434252\,\delta Y + 9.26389 \times 10^{-2}\,\delta Z  
\eea
where the change in signs is due to choosing either the positive or negative $Z$ solution.
This system has eigenvalues $(-2.99504,-2.96384,-3.11922 \times 10^{-2})$ whose real parts are all negative. This system has an average slow-roll parameter of $\epsilon = 4.96279 \times 10^{-3}$ (from both \eqref{eps3} and $\epsilon_-$ in \eqref{eps}) and anisotropy $ \Sigma/H = 3.08249 \times 10^{-6}$, which is however too large  compared to current data \cite{KK,NKY}. 
As a reference, from \eqref{u w}, we can also evaluate the constant of integration for the vector field defined in \eqref{fsol} in terms of $M_{Pl}$, $f_0$ and $\phi_0$\footnote{For example, for $(\phi_0,f_0)=(-1,1)$, we find $p_A=\pm 4.4\times 10^{-7} \, M_{Pl}^2$. Inverting \eqref{ratio1}, we can also find the value of $\dot v$ in terms of all the parameters of the model, 
$\dot v \sim ({\cal R} V 2 f^{-2} e^{2\alpha-4\sigma})^{1/2}$.}.  We can  compare this solution with the  non-derivative stable one $\lambda=0.1$, $\rho=50$, $n=0$. In that case, $\Sigma/H=4\times 10^{-4}$ and thus we clearly see that the derivative coupling decreases the level of anisotropy. 

As a second  example we take  $n=-2$, $\lambda=0.01$, and $\rho=200$. This has a stable fixed point  at: $(W,Y,Z)=(2.97549 \times 10^{-10},-9.99875 \times 10^{-3},\pm 5.17484 \times 10^{-5})$. Linearisation of the $W,Y,Z$ equations around this point gives the equations:
\bea
&&\frac{d\delta W}{d\alpha}= -2.99995\,\delta W-2.97512 \times 10^{-12} \,\delta Y \pm 3.44989 \times 10^{-5}\,\delta Z  \\
&&\frac{d\delta Y}{d\alpha}=-4.28930 \times 10^{-6}\,\delta W-3.00295\,\delta Y \pm 0.145051 \delta Z \\
&&\frac{d\delta Z}{d\alpha}=\mp 1.03586 \times 10^{-4}\,\delta W \mp 7.25192 \times10^{-2}\,\delta Y + 3.00290\times 10^{-3}\,\delta Z 
\eea
where the change in signs is due to choosing either the positive or negative $Z$ solution, respectively.
The eigenvalues for this set of equations are $(-2.99995,-2.99945,-5.00566\times 10^{-4})$. The eigenvalues' real parts are all negative and hence this fixed point is stable. Therefore, this corresponds to  a stable solution that produces anisotropy during inflation. Using \eqref{eps3}, we find the slow-roll parameter to be $\epsilon = 4.99938 \times 10^{-5}$ matching perfectly with the $\zeta_-$ solution in \eqref{zetas}, which gives the slow-roll parameter (for $\lambda,\rho,n$ given above) as $\epsilon_- = 4.99938 \times 10^{-5}$. The average anisotropy is given by $\Sigma/H= 2.97549 \times 10^{-10}$ and is thus consistent with observations \cite{KK,NKY}. In addition, $w$ is positive and real.
We can compare this solution with the non-derivative case of \cite{KSW} with a slight change in the parameters. In that case, a stable anisotropic solution can be found for $\lambda=0.01$, $\rho=500$ (and of course $n=0$), so of the same order of magnitude as the present case. For that solution, the anisotropy turns out to be $\Sigma/H=4\times 10^{-6}$ and thus in tension with current data. Again, we see that a derivative coupling helps to bring apparently excluded solutions back into agreement with observations.

\section{More general solutions}\label{Sec:4}

In the previous section we explored a suitable generalisation of the non-derivative anisotropic power-law inflation studied in \cite{KSW} where the gauge kinetic function has a monomial dependence on the inflaton's velocity. 
Our general equations, however, allow for an easy exploration of other interesting possibilities. One such possibility is the case of DBI inflation \cite{DBInfla}, where the inflaton can be identified with a D-brane position or a Wilson line. In any case, the vector field featuring on the inflationary D-brane may give rise to anisotropic solutions. 
In this model, the scalar action and  gauge kinetic function are given by  \cite{DWZ}, 
\bea\label{Pf}
P(\phi,X) = \frac{2 X \gamma }{\gamma+1} + V(\phi)\,, \qquad \qquad  f(\phi,X)= \gamma^{1/2}\,, \qquad \qquad \gamma =\frac{1}{\sqrt{1+2 h X}}\,,
\eea
where $h(\phi)$ is a function of the scalar field only (the warp factor associated with the 10-dimensional geometry where the brane is moving). 
We see that  in the non-relativistic case, when $\gamma\to 1$, the scalar field is canonically normalised and the vector field decouples from the inflaton. It is not difficult to check that power-law solutions with 
$h'/h= $ const. $V'/V=\lambda=$ const. cannot be found since the constraints $\epsilon \ll 1$, $\zeta\gg 1$, and $w>0$ \eqref{uw} cannot be simultaneously satisfied. The same happens when considering a canonically normalised inflaton \eqref{Peq} coupled disformally to the vector via \eqref{Pf}. This is consistent with the results of \cite{DisformVec} where a detailed analysis is shown.

However, motivated by the DBI anisotropic solutions found in \cite{DoKao}, we can modify slightly this Ansatz. Consider a DBI inflaton, with $P$ given as in \eqref{Pf}, and  a monomial, derivative coupling, $f$, as in \eqref{gkf}. This could correspond to a model where the inflaton and the vector  live in different D-branes. 
In \cite{DoKao} the authors found power-law anisotropic solutions with  
$h'/h= $const. which implies that $\gamma=\gamma_0 =$ const. 
Let us see this in some detail. 

\subsection{DBI inflation with monomial, derivative coupling solutions}

Considering  $h'/h=\, $const. implies an exponential form for $h$, which we take as:
\be
h(\phi)=\frac{h_0}{M_{Pl}^4} e^{\frac{\Lambda}{M_{Pl}}\phi} \, .
\ee
Taking also an exponential form for the inflaton potential as before, \eqref{Vansatz}, and power-law solutions for the scale factors and inflaton as in the  canonically-normalised case \eqref{ansatz},
we obtain the conditions:
\be
\lambda\xi = -2  \,, \qquad \rho\,\xi + 2\,\zeta + 2\,\eta + 2n =1 \,, \qquad \Lambda=-\lambda \,.
\ee
The requirement that $\Lambda=-\lambda$ is akin to setting $\gamma=\gamma_0=$ const. In terms of $h_0$, $\lambda$, and $\phi_0$; $\gamma_0$ becomes:
\be
\gamma_0=\left(1-4\frac{h_0}{\lambda^2}e^{-\lambda\phi_0}\right)^{-\frac12}\, .
\ee
 By applying the exact same procedure of balancing the amplitudes as we used for the canonically-normalised case \eqref{Ham Constraint}-\eqref{zetas}, we obtain two analogous solutions that satisfy all of the system's equations:
\be
\zeta_+=\frac{D+\sqrt{E}}{72\,n\,\lambda}\,, \qquad \zeta_-=\frac{D-\sqrt{E}}{72\,n\,\lambda}\,,
\ee
where 
\be
D=6\, \lambda+18\,n\,\lambda-36 \,n^2\,\lambda+12\,\rho+36\,n\,\rho\,,
\ee
and
\be
E=D^2-144\,n\,\lambda\left(\frac{8}{\gamma_0\,\lambda}+\lambda-2\,n\,\lambda+8\,\rho-12\,n\,\rho+12\,\frac{\rho^2}{\lambda}\right)\,.
\ee
 Anisotropic inflationary solutions can now be found for suitable choices of the parameters, as long as they satisfy the  constraints that $u,w$ (defined as before \eqref{u w}) must be real and positive, and, of course, $\epsilon=1/\zeta\ll1$.
As a concrete example, a stable solution can be found by for $\lambda=0.01,\, \rho=300,\, n=-2,\,\gamma_0=1.5$. It has anisotropy $\Sigma/H = 1.32254 \times 10^{-10}$ and slow-roll parameter $\epsilon = 3.33306 \times 10^ {-5}$, making it compatible with data. We present in the appendix the stability analysis of this solution. Comparing to the canonically-normalised case ($\gamma_0=1$) discussed above, we can see  the effect of the  DBI kinetic term, which reduces very slightly the anisotropy and slow-roll parameter: $(\Sigma/H)_{cn}=  2.97549 \times 10^{-10}, \, \epsilon_{cn} = 4.99938 \times 10^ {-5}$.

\section{Discussion}\label{conc}

We have studied anisotropic inflationary solutions where the inflaton couples to a vector field derivatively. That is, the gauge kinetic function depends both on the inflaton and its derivative, $f(\phi, X)$, with $2X=(\partial\phi)^2$. This coupling is motivated by  D-brane inflationary models, where the D-brane  features a vector on its world volume, and couples derivatively to the brane's position (or a Wilson line), the inflaton. Moreover, such couplings parameterise generic inflaton-matter couplings, which may be relevant in studies of the inflationary universe as a cosmological collider  \cite{Chen}. On the other hand, they  also  appear in the  EFT of anisotropic inflation  \cite{Hassan,Rostami}.

We started by presenting  a  general set-up, which allows for  the study of a wide range of models. We studied first an immediate generalisation of the power-law anisotropic model studied in \cite{KSW}, where the gauge kinetic function is a monomial in $X$, \eqref{gkf}, while exponential in the inflaton. We found that there are no stable inflationary solutions for a purely shift symmetric coupling (that is $f_\phi=0$). However, stable derivative anisotropic solutions arise for a large range of parameters. Interestingly, compared to the non-derivative case, the derivative anisotropic solutions have a lower level of anisotropy. We presented two illustrative examples. In the most relevant from the observational point of view, the anisotropy goes down by three orders of magnitude with respect to the non-derivative case, $\Sigma/H =4\times 10^{-6} \to 
1\times 10^{-9}$. We also found that the value of the anisotropy depends mildly on the power $n$ in \eqref{gkf}, which needs to be negative. 
We also found that the DBI generalisation of the power-law solutions in \cite{KSW} can also be extended to the derivative case. That is, derivative anisotropic DBI solutions exists, where the gauge kinetic function is a monomial in $X$ (see \eqref{gkf}). This example could correspond to a DBI inflationary  model where the inflaton and the vector field live in different D-branes. 
On the other hand, in the case where the inflaton and vector live on the same brane, the gauge kinetic function is dictated by the model and given by  \eqref{Pf}. In this case  however, the requirements of inflation, small anisotropy  and a positive vector energy density ($w>0$) are not compatible and thus there are no solutions. 

As we discussed in section \ref{Sec:3}, it is easy to understand  the decrease in the anisotropy by looking at the final value of the anisotropy, which is given by the ratio of energy density of the vector field to that of the scalar field and  given by \eqref{ratio}, thus  decreasing the level of anisotropy.
A clear follow-up is to look at how the derivative coupling affects a potential anisotropy in the power spectra. We leave this for a future publication.

\appendix

\section{Stability of the Anisotropic  DBI solutions with derivative couplings}

In this appendix we look at the stability of the anisotropic  DBI solution discussed in the main text. We define dimensionless variables analogous  to the canonically-normalised case:
\be \label{WYZgamma}
 W=\frac{\dot{\sigma}}{\dot{\alpha}}, \, \qquad Y=\frac{\gamma_0}{ M_{Pl}}\,\frac{\dot{\phi}}{\dot{\alpha}},\, \qquad Z=\frac{f(\phi,X) \,e^{-\alpha+2\,\sigma}}{M_{Pl}}\,\frac{\dot{v}}{\dot{\alpha}}.
\ee
The Hamiltonian constraint \eqref{einstein3} for this system becomes:
\be
-\frac{V}{M_{Pl}^2\,\dot \alpha^2}=3(W^2-1)+\frac{Y^2}{1+\gamma_0}+\frac{Z^2}{2}\,(1-4n)\,,
\ee
and the slow-roll parameter:
\be
\epsilon=3W^2+\frac{Y^2}{2\gamma_0}+\frac{Z^2}{3}\,(1-3n)\,.
\ee
The equations of motion in terms of \eqref{WYZgamma} become:
  \be
 \frac{dW}{d\alpha} =\frac{1}{3}Z^2\,(W+1-3 \,n\,W)+W\left(3(W^2-1)+\frac{Y^2}{2\gamma_0}\right) \,,\label{Wgammaeq2} 
 \ee
 \bea
 &&\hskip-1cm\frac{dY}{d\alpha}=F(Y,Z)\,Y\left\{3\, \gamma_0\,Y^4+4\,\gamma_0\,n\, Z^2 \left[-3 W\, (3W (4\, n-1) +4)+Z^2 \,(n\, (12\,n-7)+1)-3\right]+\right.  \nonumber \\
 && \nonumber \\
 && 2 Y^2 \left[Z^2 \left(\gamma_0^2-3 n \left(\gamma_0^2+4 n-1\right)\right)+9\, \gamma_0^2 \, W^2-9\right]+3 \lambda   Y \left(-4 n Z^2+6 W^2+Z^2-6\right) \nonumber \\ 
 && \nonumber \\
 && \left. +6 \rho\, YZ^2\,(1-4 n) +3 \gamma_0 \, \lambda \, Y^3\right\}
 \label{Ygammaeq2} 
 \eea
 \bea
 &&\hskip-1cm\frac{dZ}{d\alpha}=F(Y,Z)\,Z\left\{48\,\gamma_0\,n^3\,Z^4+2n\left[2Z^2\,\gamma_0\,(-6-6W+9W^2+Z^2)+3Y^2(-6+Z^2-\gamma_0^2Z^2) \right.\right. \nonumber \\
 && \nonumber \\
 &&\left. +3\lambda \,Y(-6+6W^2+Z^2)+3\lambda\,\gamma_0\,Y^3\right]-4n^2Z^2\left[\gamma_0 (-18+36W^2+7Z^2)+6Y(Y+\lambda)\right] \nonumber \\
 && \nonumber \\
 &&+\left. \gamma_0 \, Y^2\left[2\gamma_0(-6-6W+9W^2+Z^2)+3Y(Y-2\rho)\right]\right\}\,,\label{Zgammaeq2}
 \eea 
 where
 \be
 F(Y,Z)=\frac{1}{6\,\gamma_0\,(2\,n\,Z^2(1-4n)+\gamma_0\,Y^2)} \, .
 \ee
 This system reduces to the canonically-normalised case \eqref{Weq2}-\eqref{Zeq2} when $\gamma_0 \rightarrow 1$. This system permits stable, anisotropic solutions. As an example, a stable solution can be found by taking $\lambda=0.01,\, \rho=300,\, n=-2,\,\gamma_0=1.5$. With these parameters, we find a fixed point at $(W,Y,Z)=(1.32254 \times 10^{-10},-9.99917 \times 10^{-3},\pm 3.45004 \times 10^{-5})$. Linearisation of equations \eqref{Wgammaeq2}-\eqref{Zgammaeq2} around this fixed point yields:
\bea
&&\frac{d\delta W}{d\alpha}= -2.99997\,\delta W-8.81622 \times 10^{-13} \,\delta Y \pm 2.30003 \times 10^{-5}\,\delta Z  \\
&&\frac{d\delta Y}{d\alpha}=-1.2701 \times 10^{-6}\,\delta W-1.33371\,\delta Y \pm 6.44229 \times 10^{-2}\, \delta Z \\
&&\frac{d\delta Z}{d\alpha}=\mp 6.90183 \times 10^{-5}\,\delta W \mp 2.53082 \times10^{-2}\,\delta Y + 8.89147\times 10^{-4}\,\delta Z \,.
\eea
This fixed point has eigenvalues $(-2.99997,-1.33249,-3.33631 \times 10^{-4})$ and is therefore stable. It has anisotropy $\Sigma/H = 1.32254 \times 10^{-10}$ and slow-roll parameter $\epsilon = 3.33306 \times 10^ {-5}$, making it compatible with data.

\acknowledgements
We thank Gianmassimo Tasinato for discussions. SK was supported by IKERBASQUE, the Basque Foundation for Science and the Basque Government (IT-979-16),  
and Spanish Ministry MINECO  (FPA2015-64041-C2-1P).  IZ was partially supported by the STFC grants ST/N001419/1 and ST/L000369/1. SK would like to thank the Physics Department at Swansea University for warm hospitality and the College Research Fund for financial support.

 
\bibliography{refs}

\bibliographystyle{utphys}

\end{document}